*Thermotropic reentrant isotropy and antiferroelectricity in the ferroelectric nematic realm:*

*Comparing RM734 and DIO*

Bingchen Zhong[1], Min Shuai[1], Xi Chen[1], Vikina Martinez[1], Eva Korblova[2],

Matthew A. Glaser[1], Joseph E. Maclennan[1], David M. Walba[2], Noel A. Clark[1]*

[1]*Department of Physics and Soft Materials Research Center,*

*University of Colorado, Boulder, CO 80309, USA*

[2]*Department of Chemistry and Soft Materials Research Center,*

*University of Colorado, Boulder, CO 80309, USA*

*Abstract*

The current intense study of ferroelectric nematic liquid crystals was initiated by the observation of the same ferroelectric nematic phase in two independently discovered organic, rod-shaped, mesogenic compounds, RM734 and DIO. We recently reported that the compound RM734 also exhibits a monotropic, low-temperature, antiferroelectric phase having reentrant isotropic symmetry (the $I_A$ phase), the formation of which is facilitated to a remarkable degree by doping with small (below 1%) amounts of the ionic liquid BMIM-PF$_6$. Here we report similar phenomenology in DIO, showing that this reentrant isotropic behavior is not only a property of RM734 but is rather a more general, material-independent feature of ferroelectric nematic mesogens. We find that the reentrant isotropic phases observed in RM734 and DIO are similar but not identical, adding two new phases to the ferroelectric nematic realm. The two $I_A$ phases exhibit similar, strongly peaked, diffuse x-ray scattering in the WAXS range ($1 < q < 2$ Å$^{-1}$) indicative of a distinctive mode of short-ranged, side-by-side molecular packing. The scattering of the $I_A$ phases at small $q$ is quite different in the two materials, however, with RM734 exhibiting a strong, single, diffuse peak at $q \sim 0.08$ Å$^{-1}$ indicating mesoscale modulation with ~80 Å periodicity, and DIO a sharper diffuse peak at $q \sim 0.27$ Å$^{-1}$ ~ ($2\pi$/molecular length), with second and third harmonics, indicating that in the $I_A$ phase of DIO, short-ranged molecular positional correlation is smectic layer-like.

*INTRODUCTION*

In 2017, two groups independently reported, in addition to the typical nematic (N) phase, novel nematic phases made from strongly dipolar molecules, the "splay nematic" in the molecule RM734 [1,2,3] and a "ferroelectric-like nematic" phase in the molecule DIO [4]. These nematic phases were subsequently demonstrated to be ferroelectric in RM734 [5] and in DIO [6,7], and to be the same phase in these two materials [7,8]. This phase, the ferroelectric nematic ($N_F$), is a uniaxially symmetric, spatially homogeneous liquid having nearly saturated polar ordering of its longitudinal molecular dipoles (a polar order parameter > 0.9) [5,8,9]. DIO also exhibits an anti-ferroelectric phase, the smectic $Z_A$ ($SmZ_A$), which has smectic layers of alternating ferroelectric polarization with the director and polarization parallel to the smectic layer planes [10], and several of its close homologs have another new phase, the ferroelectric smectic A ($SmA_F$), which is also uniaxial but with polarization normal to its smectic layers [11,12]. These new phases are the gateways to the "ferroelectric nematic realm" of new liquid crystal science.

We recently reported a transition from the ferroelectric nematic liquid crystal ($N_F$) phase to a low-temperature, liquid phase having reentrant isotropic symmetry ($I_A$) in mixtures of the liquid crystal compound RM734 with small concentrations of the ionic liquids BMIM-$PF_6$ (BMIM) or EMIM-TFSI (EMIM), shown in *Fig. 1* [13]. Even a trace amount of ionic liquid dopant was found to facilitate the kinetic pathway of the transition from $N_F$ to $I_A$, suppressing crystallization and enabling the $I_A$ to form by simple cooling of the $N_F$. Unlike the $N_F$, the $I_A$ phase shows no response to applied electric field and thus appears to be nonpolar, *i.e.*, either paraelectric or antiferroelectric, with experiments to date indicating the latter. The appearance of antiferroelectric ordering adjacent in temperature to the strongly polar-ordered $N_F$ phase may seem surprising but the century-long experimental and theoretical study of ferroelectric materials, both crystals and liquid crystals, shows that ferroelectricity (F) and antiferroelectricity (A) go hand in hand, such that if one is found, the other will also be found in related materials, and that they may even appear as coexisting phases. At the root of this behavior are dipole-dipole interactions and the inherent ambivalence in how dipoles prefer to pack: dipole pairs arranged end-to-end prefer relative parallel orientation, whereas dipole pairs arranged side-by-side prefer relative antiparallel orientation. This frustration is almost a recipe for generating modulated, anisotropic, antiferroelectric phases of the SmZ$_A$ type [11], having ferroelectric stripes of uniform *P* in the end-to-end direction, and antiferroelectric ordering of adjacent stripes in the side-by-side direction. This richness of new phases and phenomena motivates the notion of a ferroelectric nematic realm, which is further broadened by the report here of the reentrance of isotropic symmetry upon cooling a highly polar and anisotropic N$_F$ liquid crystal state.

*RESULTS*

We carried out SAXS and WAXS experiments, depolarized transmission optical microscopy, and polarization current measurement of dilute mixtures of ionic liquid in DIO, focusing here on DIO/EMIM at weight% ionic liquid (IL) concentrations, $c_{IL}$, in the range (0% ≤ $c_{IL}$ ≤ 5%). The molecular species studied are shown in *Fig. 1A*. These experiments were carried out on a DIO sample synthesized as described in Ref. [8], having a phase sequence on cooling I – 173.6°C – N –84.5°C – SmZ$_A$ – 68.8°C –N$_F$ – 34°C – X. Our principal results on DIO/IL mixtures, based on x-ray and optical observations, are summarized in the phase diagram of *Fig. 1B*.

The introduction of IL can produce stunning alterations of phase behavior in mesogens of the ferroelectric nematic realm. Even a very low IL concentration ($c_{IL}$ ~ 0.01%) facilitates the appearance, upon lowering *T* below ~30⁰C, of a new, gel-like bulk phase with reentrant isotropic symmetry (*Fig. 1B*), which we term the I$_A^{lam}$, where the subscript A signifies "antiferroelectric," distinguishing this phase from that of the usual dielectric isotropic phase (I) found at high temperature, and the superscript "lam" (for "lamellar"), distinguishing this isotropic phase from the I$_A$, the reentrant isotropic phase observed in RM734 (*Fig. 1C*). Remarkably, even with such low concentrations of IL, the first-order thermotropic transition upon cooling from the N$_F$ to the I$_A^{lam}$ phase converts the entire sample volume to I$_A^{lam}$, suggesting that the I$_A^{lam}$ phase is intrinsic to the DIO host, with the IL dopant serving to create a kinetic pathway for its nucleation. As in undoped DIO, the SmZ$_A$ phase appears between the N and N$_F$ phases, becoming more stable relative to these phases with increasing IL concentration. The onset of phase separation of the IL component is observed when the dopant concentration $c_{IL} \gtrsim$ 1% (shaded region). Here we focus on the mesophase behavior at smaller IL concentrations.

In our previous experiments with RM734 as the host [13], the dopant IL was BMIM. In the present case, the DIO host was doped with EMIM but we have found that BMIM and EMIM behave very similarly in both materials, with both ILs facilitating the kinetic pathway to a reentrant isotropic phase, the structure of which is a property of the host $N_F$ and does not depend on the IL concentration (see *Fig. 2B,C* and Fig. 2B of Ref. [13]) or on which ionic dopant is used (see Fig. 7 of Ref. [13]).

*X-ray scattering* – Powder-average SAXS and WAXS scans of temperature-controlled samples in 0.7 mm to 1 mm diameter thin-wall capillaries were obtained during slow cooling, with results shown in *Figs. 2,4-6*.

*Temperature dependence with $c_{IL}$ = 1%* – A selection of x-ray scans at temperatures in the N, SmZ$_A$, N$_F$, I$_A^{lam}$, and X phases in the $c_{IL}$ = 1% mixture are shown in *Fig. 2A*. Complete cooling sequences for this and the other $c_{IL}$ values are provided in the Supplementary Information (SI). The higher-temperature, anisotropic, N, SmZ$_A$, and N$_F$ phases each exhibit a broad, diffuse WAXS peak at $q$ ~ 1.5 Å$^{-1}$, the powder average of the equatorial diffuse peak observed in the WAXS of magnetic field-aligned N and N$_F$ DIO samples due to the side-by-side positional correlation arising from steric repulsion of the molecules [1,8]. The transition to the I$_A^{lam}$ phase is marked by significant changes in this scattering, with the broad WAXS peak breaking up into several much narrower diffuse peaks with scattering wavevectors $q$ in the range (0.8 Å$^{-1}$ < $q$ < 2.2Å$^{-1}$), marking the development of specific, longer-ranged, side-by-side intermolecular positional correlations in the I$_A^{lam}$ phase.

*Dopant concentration dependence at $T = 20^0C$* –The x-ray scattering $I(q)$ of the I$_A^{lam}$ phase at $T = 20^0C$ at several different concentrations, $c_{IL}$, of EMIM in DIO is shown in *Fig. 2B*. It is evident from these scans that the structure of the x-ray peaks in the I$_A^{lam}$ phase varies only very weakly with concentration for $c_{IL}$ in the range (0.01% ≤ $c_{IL}$ ≤ 5%), in both the SAXS and WAXS regimes. This, and the fact that the I$_A^{lam}$ phase is observed at very low dopant concentrations, suggests that the causative molecular organization of the I$_A^{lam}$ phase is intrinsic to the DIO host.

*Comparison of reentrant isotropic scattering of DIO with that of RM734* – The reentrant isotropic structure functions $I(q)$ of pure RM734 and DIO at $T = 20^0$C are compared directly in *Fig. 3*. Remarkably, the WAXS scattering of the I$_A^{lam}$ phase of DIO bears a strong resemblance to that observed in the I$_A$ phase of RM734, with two prominent diffuse peaks in each scan, a similarity that suggests that the existence of the I$_A$ is a general phenomenon of ferroelectric nematic mesogens, and not just a peculiarity of RM734. The scattering at shorter wavevectors ($q \lesssim 0.8$ Å$^{-1}$) in the two materials is, however, quite different. The single, small-angle diffuse peak of RM734 is located at $q_M \approx 0.080$ Å$^{-1}$ but DIO exhibits a sequence of three, narrower, diffuse peaks, at $q_M \approx 0.27$, 0.55, and 0.83 Å$^{-1}$, a 1:2:3 harmonic series indicating strong local lamellar order with a layer spacing $d_M \sim 27$ Å, approximately the molecular length of DIO. The fundamental is at the same wavevector as the strong, diffuse scattering feature along $q_z$, where $z$ is parallel to the director, observed in all of the LC phases of DIO [8,10] and close to that of the smectic A$_F$ phase observed in binary mixtures of DIO with the mesogen 2N [11], and of close molecular analogs of DIO [12,14]. Thus, the tendency of DIO to form locally layered structures with molecular monolayer spacing is also manifest in its reentrant isotropic phase as short-ranged, molecular monolayer lamellar ordering. We consequently term the reentrant isotropic in DIO the I$_A^{lam}$ phase.

*50% RM734 / 50% DIO* – In order to further explore reentrant isotropy as a phenomenon of the ferroelectric nematic realm, we studied the reentrant isotropic phase behavior of a 50% RM734/50% DIO mixture, with results shown in *Fig. 4*. Optical microscopy of capillaries of this mixture, shown in *Fig. 4B*, provides evidence for the coexistence of two isotropic phases. X-ray scattering of a sample volume containing many dispersed droplets of one phase in the other exhibits the principal features of the individual RM734 and DIO scans in *Fig. 3* but cannot be modeled as a linear combination of them (recalling that $I(q)$ of the reentrant isotropic in both doped RM734 and doped DIO does not depend strongly on $c_{IL}$). Specifically: (*i*) The main peaks in the WAXS region are of comparable width and intensity to those of RM734 and DIO but are at different $q$-values; (*ii*) The DIO-like SAXS peaks are weaker relative to the WAXS peaks for both the RM734 and DIO contributions to the black curve in *Fig. 4A*; (*iii*) The peak corresponding to small-$q$ lamellar scattering in doped DIO is shifted in wavevector from a fundamental wavevector $q_M$ = 0.276 Å$^{-1}$ (white lines) to $q_M$ =0.297 Å$^{-1}$ (green lines) in the 50/50 mixture (*Fig. 4A*); (*iv*) The diffuse SAXS peak similar to that in RM734 is at $q_M$ = 0.125 Å$^{-1}$. These differences indicate that each of the coexisting I$_A$ phases in the 50/50 mixture must have some minority fraction of either RM734 or DIO in its individual composition.

*Optical textures and electro-optics of DIO / EMIM mixtures* – Polarized light microscopy observations of the mixtures were made in cells with the LC between a pair of glass plates spaced 3.5 μm apart. One plate was coated with a pair of ITO electrodes separated by a 1.04 mm wide gap used to apply an electric field, *E*, largely parallel to the sample plane. These cells were filled and studied only at temperatures below 120°C, as the components thermally degraded at higher temperatures, resulting in irreversible changes in phase behavior. The plates were treated with polyimide layers with antiparallel buffing along a direction 3° from being parallel to the electrode edges. In the N phase, this preparation produces uniform, planar alignment of the director *n(r)*, the local mean molecular long-axis and the optic axis, parallel to the glass and 3° from normal to the field. Upon cooling into the N$_F$ phase, the *n(r)*-*P(r)* couple transitions to a π-twisted geometry, a result of the antiparallel orientation of the ferroelectric polarization, *P(r)*, on the two surfaces, as previously detailed [9]. The angular offset of the rubbing direction ensures a well-defined initial reorientation direction of *n(r)* with application of a given sign of electric field.

Cooling a $c_{IL}$ = 0.2% DIO/EMIM mixture from the N to the N$_F$ phase at T = 30°C results in the formation of uniformly twisted N$_F$ domains, shown in *Fig. 5A*. Further cooling induces the N$_F$ to I$_A$$^{lam}$ phase transition at T ≈ 27°C, where small, extinguishing I$_A$$^{lam}$ domains nucleate via a first-order transition, as seen in *Fig. 5B*. These I$_A$$^{lam}$ domains, which are dark between crossed polarizer and analyzer at all sample orientations, eventually grow to cover the entire sample area (*Fig 5C*). The extinction of the final I$_A$$^{lam}$ dark state between crossed polarizers is comparable to that of the high temperature I phase. In all cells, and at all $c_{IL}$ concentrations, we observe a patchy texture of weakly transmitting areas of low remnant birefringence, apparently due to sub-10 nm thick birefringent layers on the glass surfaces in which there may be induced nematic order. Application of in-plane electric fields of a few V/mm that easily reorient the bulk director/polarization couple in the N$_F$ phase to be nearly along the field [5] has no visible effect on the dark $I_A^{lam}$ domains, suggesting that the $I_A^{lam}$ structure may be antiferroelectric.

*SmZ$_A$ phase* – Undoped DIO exhibits a phase between the N and N$_F$, the smectic Z$_A$, which we have shown previously to be lamellar, having an antiferroelectric array of layers with the polarization parallel to the layer planes [10]. The SmZ$_A$ layering in undoped DIO appears in x-ray diffraction as very weak peaks in the SAXS range which to date have required a synchrotron source to observe [10,11]. Otherwise, the x-ray scattering from the SmZ$_A$ in the SAXS/WAXS ranges obtained using our in-house diffractometer does not differ substantially from that of its neighboring N and N$_F$ phases, so is not useful for phase identification. However, the lamellar nature of the SmZ$_A$ phase is also manifest in a host of characteristic textural features previously observed in the study of smectics, which, on their own, enabled identification of this intermediate phase as a smectic in undoped DIO, using the cell geometries and characterization methods, including electric field application and temperature scanning, described in Ref. [10].

The same experimental methods were applied to the DIO/EMIM mixtures, leading to the phase diagram of *Fig. 1*, in which the phase between N and N$_F$ is identified as SmZ$_A$. As noted, the x-ray scattering from this phase in the doped sample, exemplified for $c_{IL}$ = 1% in *Fig. 2A*, is quite similar to that of its neighboring N and N$_F$ phases. However, the visual signatures of the SmZ$_A$ layer ordering in the cell textures are quite unmistakable across the concentration range shown in *Fig. 1*. An example of this is shown in *Fig. 6*, in which comparable arrays of zig-zag smectic layering defects are shown in $c_{IL}$ = 0 and $c_{IL}$ = 1% DIO/EMIM samples. This requires nearly identical birefringence, the same alignment of *n* parallel to the buffing, the same alignment of layer normal $q_M$ nearly parallel to the plates, and the same layer organization within the zig-zag walls, defects that mediate reversal of the smectic layer chevron direction. Similarities such as these lead to the identification of this intermediate phase in DIO/EMIM mixtures as SmZ$_A$.

Significantly, as indicated in *Fig. 1*, while undoped RM734 has no SmZ$_A$ phase, the addition of a small amount of IL dopant ($c_{IL}$ = 1%) is sufficient to induce a 20°C-wide, SmZ$_A$-like phase between the N and N$_F$ phases. This phase is antiferroelectric, has the layer normal $q_M$ oriented perpendicular to the director *n*, and exhibits zig-zag walls as shown in Fig. 5 of Ref. [13], all characteristic features of the smectic $Z_A$ phase but with layers that are several μm thick, much larger than the 8 nm layer spacing of the SmZ$_A$ phase in DIO.

*Ferroelectric nematic realm* – The notion of the "ferroelectric nematic realm" as a field of activity and interest in LCs was motivated by our observation of the ideal binary miscibility of RM734 and DIO in the ferroelectric nematic (N$_F$) phase [8]. These two molecular species exhibit the same ferroelectric nematic phase and possess some required but yet to be definitively identified commonality that enables them to behave as equivalent participants in an N$_F$ fluid mixture. This transforms N$_F$ fluid formation into a mechanism of selection, which, given the late date of discovery of the N$_F$, is apparently rare among mesogens. In the case of the N$_F$, this selection has led, remarkably, to the observation of other new phases like the chiral N$_F$, the smectic Z$_A$, and the smectic A$_F$, which are related to the N$_F$ by having polar ordering of similar magnitude on some length scale.

These connections define the ferroelectric nematic realm, a field which is significantly enriched and enhanced by the results presented here: the surprising appearance of reentrant isotropy in RM734 [13] is now found to be a trick that DIO also performs, albeit with its own twist. The N$_F$ phase, with its saturated quadrupolar order parameter and nearly perfect polar order even at elevated temperature, can perhaps be considered the most-ordered nematic phase, combining nearly perfect long-range polar orientational order and short-range positional disorder. The I$_A$-type phases appear to be an exchange of this well-ordered N$_F$ state for one having isotropic, long-range orientational order combined with exceptionally robust, short-range molecular positional correlations (*Fig 2A*). The thermodynamic availability of the I$_A$-type phases upon cooling points to internal energetic frustration in the N$_F$ as the principal driving force for this transition.

The position of the small-angle scattering peak in DIO ($2\pi/q_M = d_M \approx 28$ Å) is indicative of local monolayer lamellar electron-density modulation which, given the isotropy and transparency of the I$_A^{lam}$ observed optically, would be some form of short-ranged smectic A$_F$ ordering [11]. However, this peak is diffuse, of half-width at half-maximum (HWHM) $\delta q \sim 0.023$ Å$^{-1}$, indicating an exponential decay of layer coherence with a characteristic distance of ~42 Å, *i.e.*, local structures of just a few layers. The lack of dependence of the I$_A^{lam}$ x-ray peak structure on ionic dopant concentration seen in *Fig. 2B*, in both the SAXS and WAXS wavevector regimes, indicates that the causative molecular organization of the $I_A^{lam}$ phase is predominantly a property of the DIO host.

Insight into the local structure of the DIO host can be gained from previous studies of systems which have sets of distinct WAXS peaks similar to those found in DIO and RM734, as well as smectic-like lamellar reflections at lower $q$ as seen in DIO, and which have been structurally characterized by freeze-fracture transmission electron microscopy (FFTEM). A review of the variety of mesogenic systems exhibiting reentrant isotropic phases [13] uncovered several such systems, including the 50 wt% mixture of the rod-shaped mesogen 8CB with the material W624 (compound *2b* in Ref. [15]), where a thermotropic Smectic C – to – Isotropic* dimorphism was observed [16]. X-ray scattering (*Fig. 7C*) and extensive FFTEM visualization of the local structure (*Fig. 7D*) in the spontaneously chiral Isotropic* (dark conglomerate) phase [16] showed it to be lamellar with strong side-by-side molecular positional correlations, its x-ray scattering bearing a strong resemblance to that of DIO and RM734 shown in *Fig. 7A,* and having a strong tendency for local saddle-splay layer curvature (see *Fig. 7D*), the latter driving the assembly of gyroid-like branched arrays of filaments of nested cylindrical layers. A similar low-$T$ isotropic phase was found also in the single-component bent-core molecule 12-OPIMB [16]. The local nested-cylinder layer structure of these phases, exemplified in the W624/8CB mixture, features director splay everywhere, a mechanism which has also been proposed to stabilize phases in the ferroelectric nematic realm [17,2]. A third, quite different low-$T$ isotropic system that gives x-ray scattering more like the $I_A^{lam}$ phase of DIO shown *Fig. 7B* is that of a small, azo-based molecule (W470), which has a chiral smectic C phase between a high-temperature isotropic phase and a low-temperature reentrant isotropic phase, with FFTEM showing the latter to form side-by-side, linear aggregates and gels when diluted with solvent [18]. However, the bulk structure of the reentrant isotropic phase of neat W470 has not been established.

### *MATERIALS AND METHODS*

The mixtures were studied using standard liquid crystal phase analysis techniques previously described [5,8,9], including depolarized transmission optical microscopic observation of LC textures and response to electric field, x-ray scattering (SAXS and WAXS), and techniques for measuring polarization and determining electro-optic response.

*Materials* – DIO was synthesized by the Walba group as described in [10] (DIO phase sequence: I – 173.6°C – N – 84.5°C – SmZ$_A$ – 68.8°C – N$_F$ – 34°C – X). EMIM-TFSI {1-Ethyl-3-methylimidazolium bis(trifluoromethylsulfonyl)imide} was obtained from Millipore/Sigma and used without further purification.

*Methods – Obtaining the I$_A^{lam}$ phase in undoped DIO or DIO/IL mixtures* – Samples of DIO and its mixtures with ionic liquid were heated into the N phase at 120°C before they were loaded into capillaries or liquid crystal cells. After filling, the x-ray capillaries were quenched on a flat metal surface at $T$ = -19°C, by which means the entire volume of the doped LC rapidly transitioned to the optically transparent reentrant isotropic phase. No annealing was required. The capillaries were then heated back to room temperature to carry out x-ray diffraction in the dark phase, after which they were heated to the uniaxial nematic phase and cooled at -0.5°C per minute to selected temperatures where x-ray diffraction measurements of the N and SmZ$_A$ phases were carried out. In the EMIM/DIO mixtures, crystals of DIO typically started to appear when the sample was cooled to between 50°C and 40°C, making it challenging to obtain diffraction images of the N$_F$ phase. The capillary could be reheated to the nematic and quenched to the I$_A$ phase as many times as desired.

The phase transition temperatures in cells and capillaries were determined using depolarized transmission optical microscopy while cooling the samples at -0.5°C per minute. DIO crystals typically started to nucleate and grow at around 40°C in such temperature scans. In order to favor formation of the I$_A^{lam}$ phase rather than the crystal, it is necessary to cool the samples more quickly, which prevents the crystal phase from growing to cover the whole volume before the transition to the I$_A^{lam}$ phase can take place. For example, crystallization could be suppressed by cooling at -3°C per minute, by which means a cell with the I$_A$ phase filling the entire volume could be obtained. In order to measure the precise temperature of the N$_F$ – I$_A$ transition, the cell was first quenched to 30°C and then cooled at -0.5°C per minute to the transition at about 27°C.

The I$_A^{lam}$ phase in the DIO/EMIM mixtures is apparently less thermodynamically stable than the crystal. The crystal phase was typically observed to start nucleating following overnight storage at 20°C. However, the I$_A^{lam}$ phase was stable for at least five days if the sample was held at 9°C. Upon heating at 0.5°C per minute, the I$_A^{lam}$ phase transitioned to the N$_F$ phase at $T \sim 45$°C, after which the sample rapidly crystallized.

*X-ray scattering* – For SAXS and WAXS experiments, LC samples were filled into thin-wall capillaries 0.7 to 1 mm in diameter. Data presented here are powder averages obtained on cooling using a Forvis microfocus SAXS/WAXS system with a photon energy of CuK$_\alpha$ 8.04 keV (wavelength = 1.54 Å). Each scan took ~1 hr at a given temperature.

*Polarized light microscopy* – Optical microscopy of LC cells viewed in transmission between crossed polarizer and analyzer, with such cells having the LC between uniformly spaced, surface-treated glass plates, provides key evidence for the macroscopic polar ordering, uniaxial optical textures, and fluid layer structure in LC phases and enables direct visualization of the director field, *n*(*r*), and, apart from its sign, of *P*(*r*).

*Electro-optics* – For making electro-optical measurements, the mixtures were filled into planar-aligned, in-plane switching test cells (Instec, Inc.) with unidirectionally buffed alignment layers arranged antiparallel on the two plates, which were uniformly spaced 3.5 μm apart. In-plane ITO electrodes on one of the plates were spaced by a 1 mm wide gap and the buffing was along a direction rotated 3° from parallel to the electrode edges. Such surfaces give a quadrupolar alignment of the N and SmZ$_A$ directors along the buffing axis and polar alignment of the N$_F$ at each plate, leading to a director/polarization field in the N$_F$ phase that is parallel to the plates and has a π twist between the cell surfaces [9].

*ACKNOWLEDGEMENTS*

This work was supported by NSF Condensed Matter Physics Grants DMR 1710711 and DMR 2005170, by Materials Research Science and Engineering Center (MRSEC) Grant DMR 1420736, by the State of Colorado OEDIT Grant APP-354288, and by a grant from Polaris Electro-Optics. X-ray experiments were performed in the Materials Research X-Ray Diffraction Facility at the University of Colorado Boulder (RRID: SCR_019304), with instrumentation supported by NSF MRSEC grant DMR-1420736.

## FIGURES

*Figure 1*: (**A**) Structures of the polar mesogens RM734 and DIO and the ionic liquids BMIM and EMIM. (**B**) Phase diagram of DIO and DIO/EMIM mixtures showing the $I_A^{lam}$, a new phase of the ferroelectric nematic realm. Pure DIO exhibits the following phase sequence on cooling from 200°C: high-temperature dielectric isotropic (I); dielectric nematic (N); antiferroelectric smectic $Z_A$ (Sm$Z_A$); ferroelectric nematic ($N_F$); and, depending on the cooling process, crystal (X) or low-temperature antiferroelectric isotropic ($I_A^{lam}$). The phase diagram of the DIO/EMIM mixtures was determined by x-ray scattering and polarized light microscopy experiments. The times and temperatures required for crystallization were highly dependent on dopant concentration $c_{IL}$ and on sample geometry, with longer times and lower temperatures required as $c_{IL}$ was increased. The shaded region indicates where, at higher dopant concentration, phase separation of EMIM is observed. (**C**) RM734 /BMIM phase diagram for comparison (reprinted with permission from Ref. [13]). The doped mixtures now both exhibit a modulated, antiferroelectric phase identified as Sm$Z_A$. As in the DIO mixtures, phase separation occurs at higher

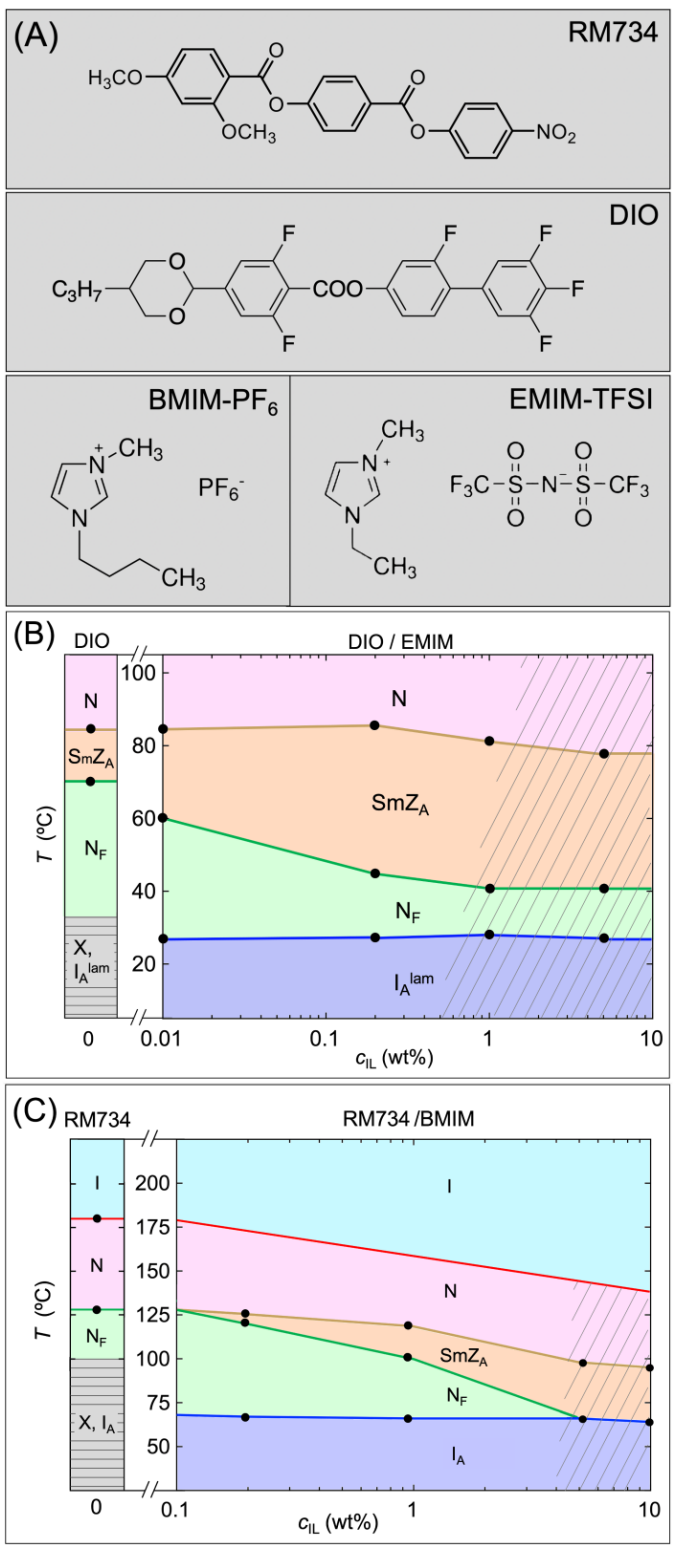

*-13-*

dopant concentrations (shaded region). Exposing mixtures to temperatures higher than $T \sim 150°C$ produces irreversible changes in phase behavior, so the I – N transition temperature in the mixtures was evaluated only approximately.

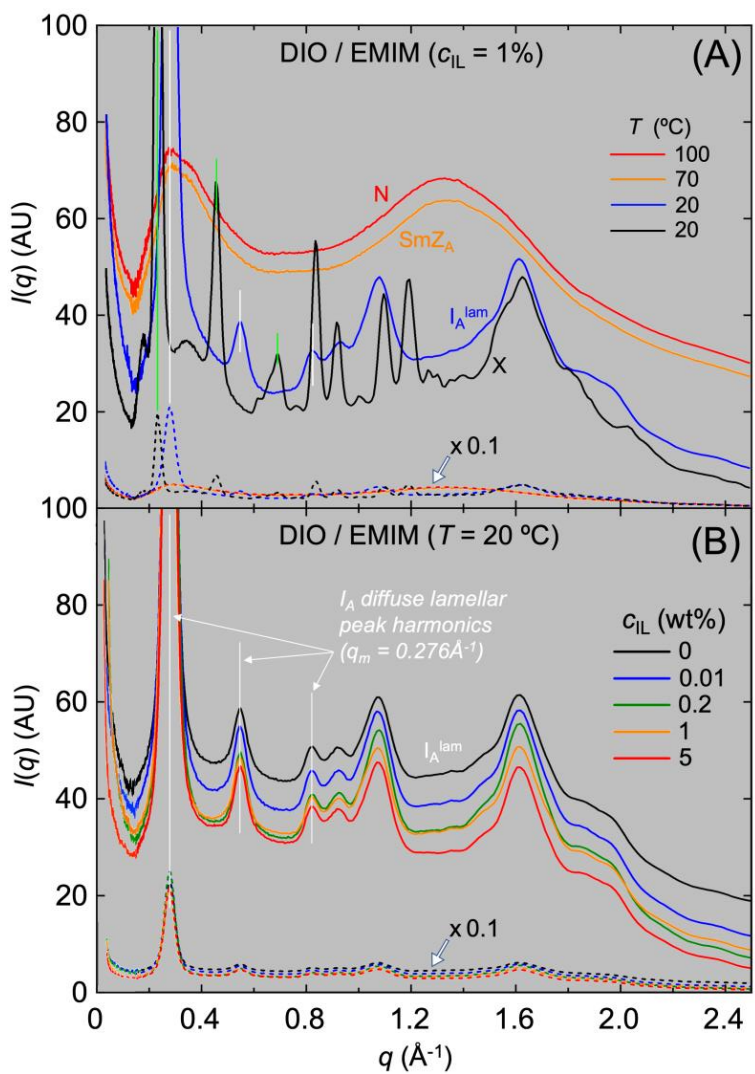

*Figure 2*: WAXS scans obtained during slow cooling of DIO/EMIM mixtures. (*A*) Scans obtained on a $c_{IL}$ = 1wt% DIO/EMIM mixture at selected temperatures show the structure functions $I(q)$ typical of each of the N, SmZ$_A$, I$_A^{lam}$, and X phases. The complete set of scans vs. $T$ is shown in the Supplementary Information. The I$_A^{lam}$ phase exhibits a diffuse peak at $q_M$ = 0.2764 Å$^{-1}$ and its first two harmonics (white lines), indicating a lamellar periodicity corresponding to the DIO molecular length, with end-to-end periodicity $d_M \sim$ 28 Å. The distinctive pattern of diffuse peaks in the WAXS $q$-range indicates side-by side molecular packing which is very different from that of the crystal (X) phase. (*B*) Scans of the x-ray structure function $I(q)$ of the I$_A^{lam}$ phase versus IL concentration $c_{IL}$. The dependence on $c_{IL}$ is very weak, showing that $I(q)$ of the I$_A^{lam}$ phase is a property of the undoped DIO host. The dashed curves are the solid curves reduced by a factor of 10.

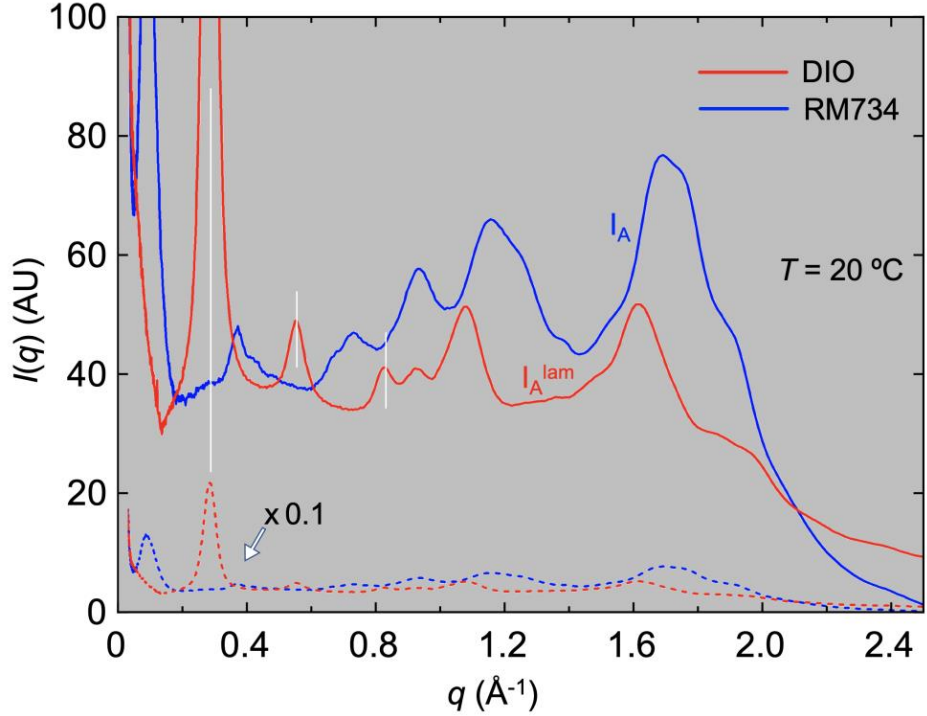

*Figure 3*: Comparison of x-ray diffraction scans of the reentrant isotropic phases of pure RM734 and DIO. The DIO $I_A^{lam}$ phase exhibits a series of three diffuse peaks at $q_M$ = 0.2764 Å$^{-1}$ and its first two harmonics (white lines), indicating a lamellar periodicity corresponding to the DIO molecular length (hence the name $I_A^{lam}$), and a distinctive pattern of diffuse peaks in the WAXS $q$-range from side-by-side molecular packing. As shown in Fig. 2B of Ref. [13], and in *Fig. 2B* above, the form of the scattering of the reentrant isotropic phases in RM734 and in DIO does not depend significantly on ionic dopant concentration.

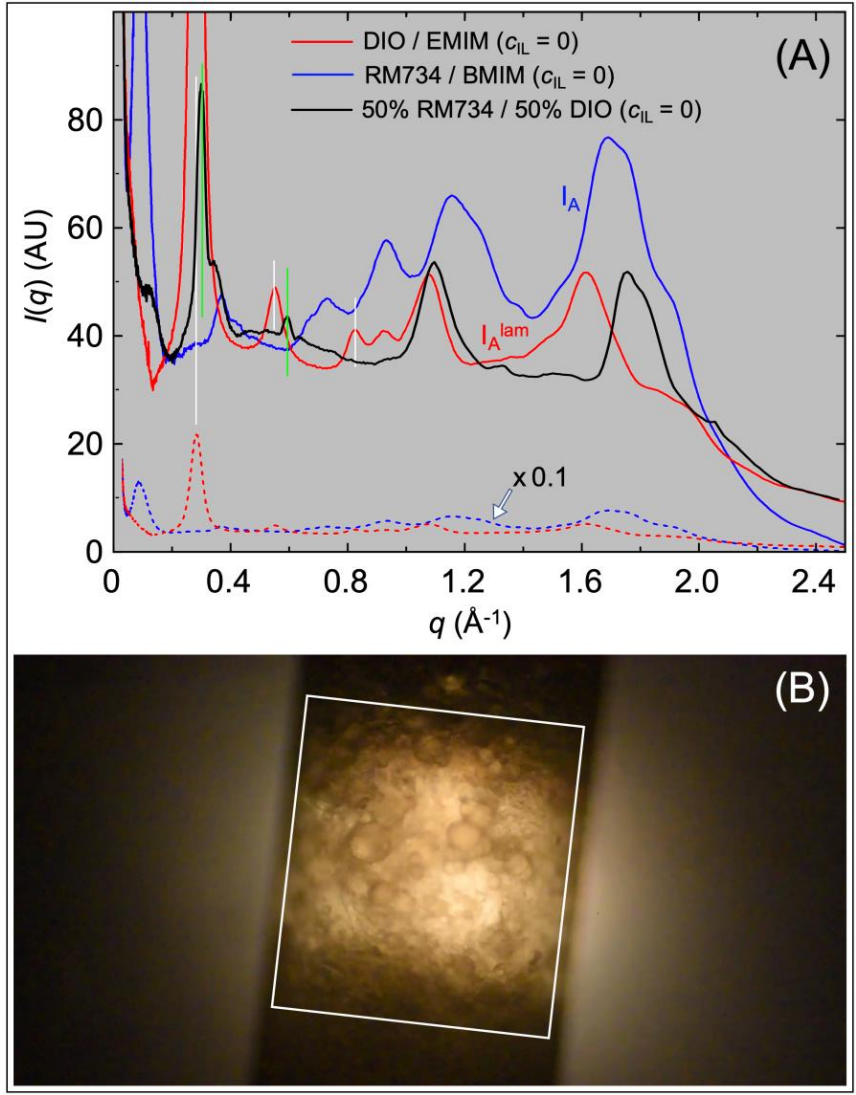

*Figure 4*: (*A*) Comparison of x-ray diffraction scans of the reentrant isotropic phases of: DIO/EMIM ($c_{IL}$ = 0) (*red*); and RM734/BMIM ($c_{IL}$ = 0) (*blue*); with 50% RM734/50% DIO ($c_{IL}$ = 0) (*black*). The x-ray scattering from the reentrant isotropic phases in the 50/50 mixture exhibit the principal features of those of the scans of doped RM734 and DIO, but with the following notable differences: (*i*) Prominent WAXS peaks similar to those observed in the individual doped materials are present in the 50/50 mixture but the peak positions and structures are not describable as superpositions of the RM734 and DIO WAXS peaks. Also, the peak corresponding to small-$q$, lamellar scattering in doped DIO is shifted in wavevector from $q_M$ = 0.276 Å$^{-1}$ to 0.297 Å$^{-1}$ in the 50/50 mixture. (*ii*) The SAXS peaks of doped RM734 and DIO are also present in the 50/50 mixture, but at much lower intensity relative to the WAXS than in the RM734 and DIO scans, reduced by factors of between five and ten, as seen by comparing the solid black curve with the red and blue ones. These differences indicate that each of the coexisting I$_A$ phases in the 50/50 mixture has a substantial minority fraction in its composition. The local molecular packing thus appears to be maintained in the coexisting phases, but at larger length scales the ordering is disrupted.

(**B**) Transmitted light image of a 50%RM734/50% DIO mixture in a 0.7 mm diameter x-ray capillary cooled from the N phase to 50°C, showing the phase separation of RM734- and DIO-rich globules, as well as the x-ray beam dimensions (white square). These globular domains are maintained respectively as distinct $I_A$ and $I_A^{lam}$ phases at room temperature.

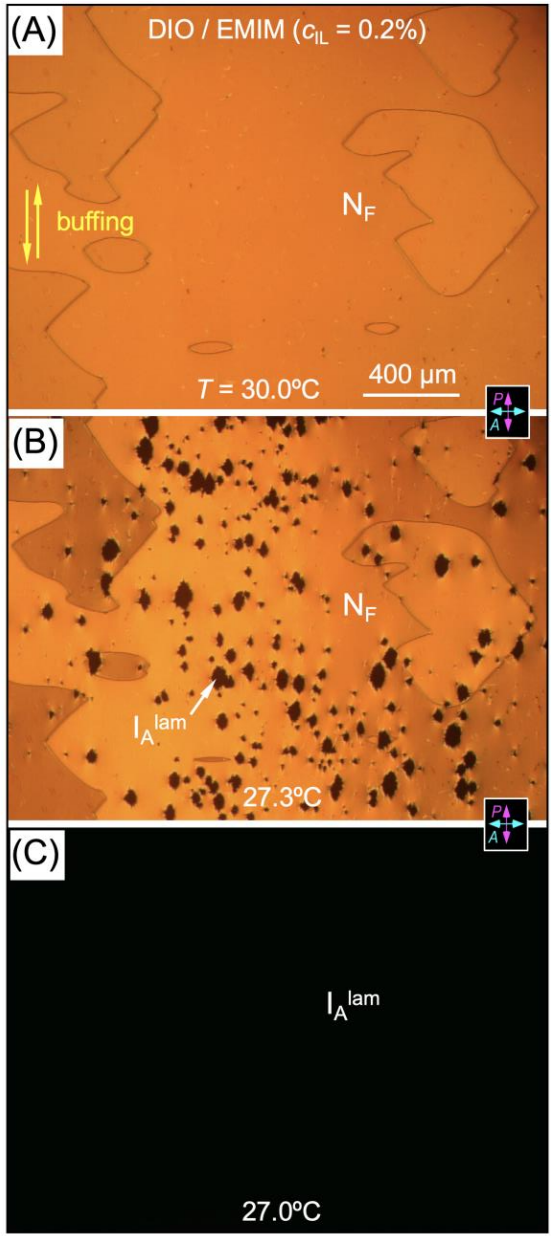

*Figure 5*: (**A**) Depolarized transmission optical microscopy image of a 3.5 μm thick DIO/EMIM ($c_{IL}$ = 0.2%) mixture in the $N_F$ phase between glass plates with antiparallel-rubbed alignment layers. This surface treatment imposes a director structure in the $N_F$ phase that is uniformly twisted by π from one plate to the other [9], producing an orange birefringence color between crossed polarizer and analyzer. Domains with twist of opposite handedness have slightly different hues. (**B,C**) Following continuous slow cooling, at around 27°C, dark, optically isotropic domains of the $I_A^{lam}$ phase nucleate and grow, eventually covering the entire cell area. At low temperatures, the $I_A^{lam}$ region has an extinction coefficient comparable to that of the high-temperature isotropic phase, and of air bubbles in the same cell [13].

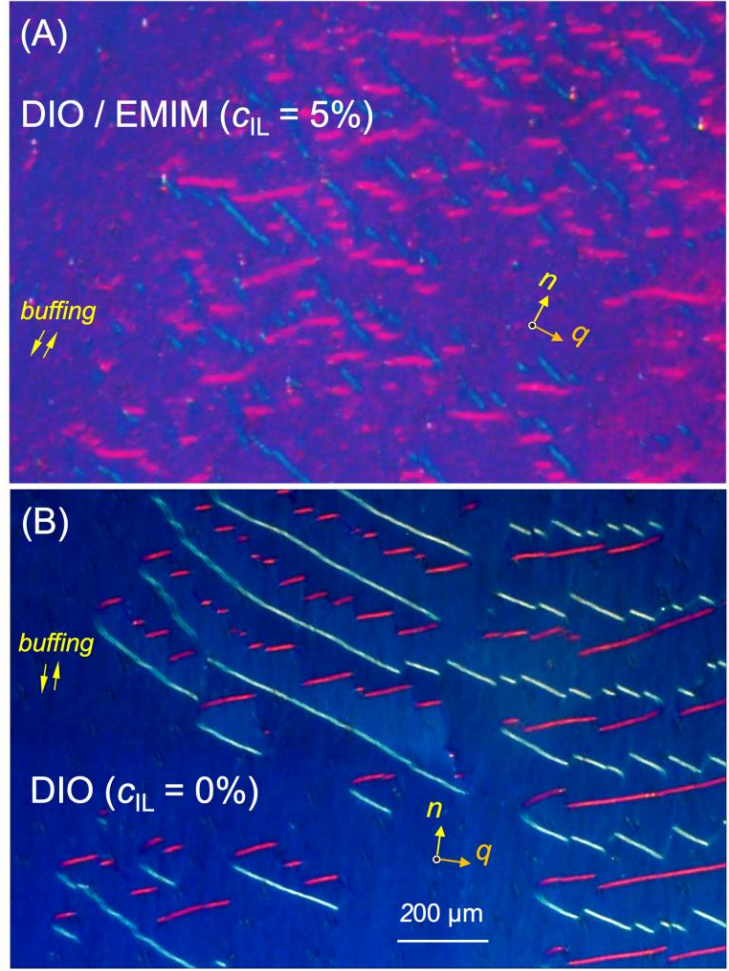

*Figure 6*: Comparison of the textures of the modulated phase intermediate between the N and N$_F$ phases in a DIO/EMIM mixture and of the SmZ$_A$ phase in a pure DIO sample: (*A*) DIO/EMIM ($c_{IL}$ = 1%) mixture; (*B*) undoped DIO ($c_{IL}$ = 0%). The samples are contained in $d$ = 3.5 μm cells that have antiparallel-buffed, polyimide-coated plates. The phases exhibit very similar characteristics: similar birefringence; alignment of *n* parallel to the buffing; layer normal orientation $q_M$ nearly parallel to the bounding plates; and chevron-layering reversal defects (zig-zag walls) (cyan and magenta in these images). These similarities lead to the identification of this intermediate phase in DIO/EMIM mixtures as SmZ$_A$. Detailed discussion of the zig-zag wall structure can be found in Fig. 4 of Ref. [10]. (*B*) is reprinted from Ref. [10] with permission.

*Figure 7*: Comparison of x-ray diffraction from the reentrant isotropic phases of RM734 and DIO with the scattering from other mesogenic systems that exhibit a reentrant isotropic phase and have similar x-ray diffraction signatures. (*A*) Full x-ray scans (SAXS+WAXS) of the $I_A$ phase in RM734 and the $I_A^{lam}$ phase in DIO, with the diffuse lamellar peaks in DIO indicated by white lines. (*B*) SAXS+WAXS scans of the reentrant isotropic phase of the single-component, chiral calamitic mesogen W470, from Ref. [18]. The molecule gelates when dissolved, forming networks of nanoscale filaments, but the structure of the reentrant isotropic phase has not been visualized. (*C*) SAXS+WAXS scans of the reentrant isotropic phase of a 50/50 mixture of the bent-core mesogen W624 and 8CB. The white lines indicating the SAXS diffuse peak wavevectors are at *q*-values with the ratio 1:2, indicating that the local structure is lamellar (smectic-like), as in DIO. (*D*) FFTEM image of the local structure of the reentrant isotropic phase of W624/8CB, showing a network of nested smectic cylinders which form a disordered array of multiply connected filaments (locally a gyroid sponge phase). This is a possible model of the DIO $I_A^{lam}$ phase structure. The data in (*B)* are reproduced from Ref. [18] with permission. The data in (C, *D*) are reproduced from Ref. [16] with permission.

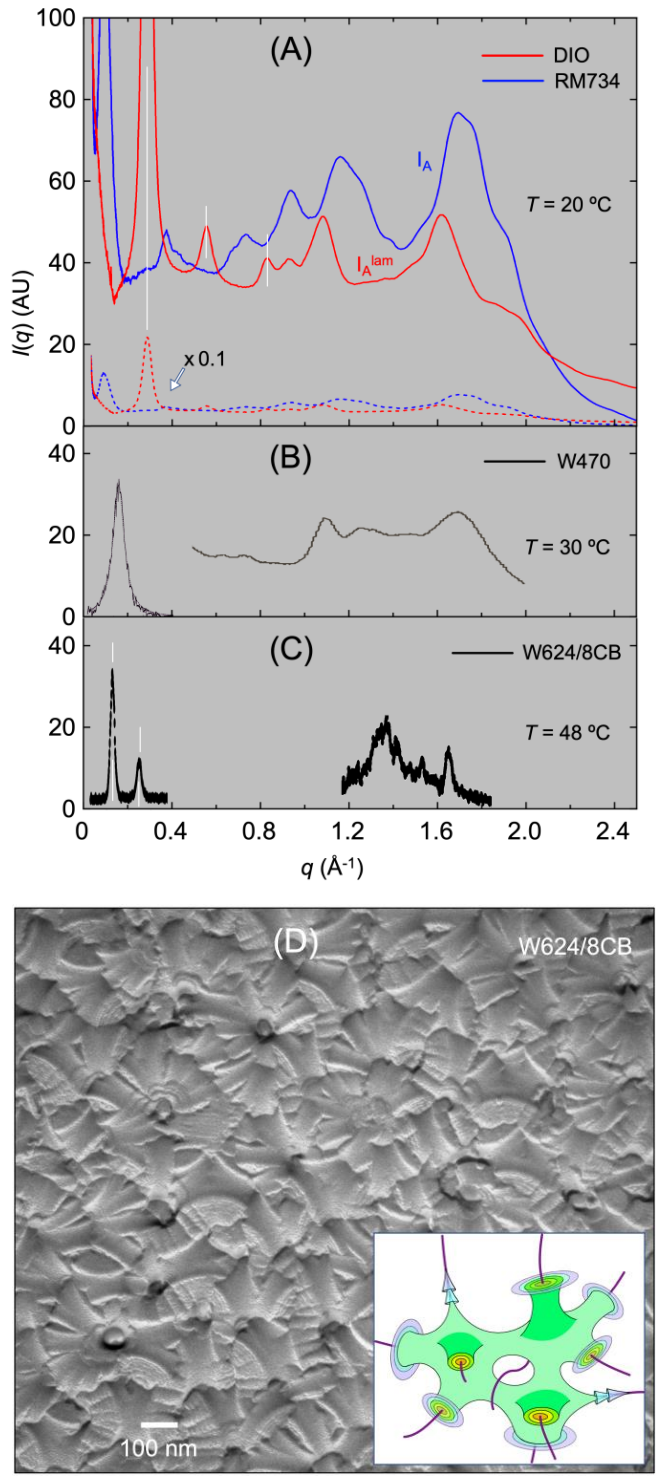

*Thermotropic reentrant isotropy and antiferroelectricity in the ferroelectric nematic realm: Comparing RM734 and DIO*

Bingchen Zhong[1], Min Shuai[1], Xi Chen[1], Vikina Martinez[1], Eva Korblova[2], Matthew A. Glaser[1], Joseph E. Maclennan[1], David M. Walba[2], Noel A. Clark[1]*

[1]*Department of Physics and Soft Materials Research Center,*
*University of Colorado, Boulder, CO 80309, USA*

[2]*Department of Chemistry and Soft Materials Research Center,*
*University of Colorado, Boulder, CO 80309, USA*

*Abstract*

The current intense study of ferroelectric nematic liquid crystals was initiated by the observation of the same ferroelectric nematic phase in two independently discovered organic, rod-shaped, mesogenic compounds, RM734 and DIO. We recently reported that the compound RM734 also exhibits a monotropic, low-temperature, antiferroelectric phase having reentrant isotropic symmetry (the I$_A$ phase), the formation of which is facilitated to a remarkable degree by doping with small (below 1%) amounts of the ionic liquid BMIM-PF$_6$. Here we report similar phenomenology in DIO, showing that this reentrant isotropic behavior is not only a property of RM734 but is rather a more general, material-independent feature of ferroelectric nematic mesogens. We find that the reentrant isotropic phases observed in RM734 and DIO are similar but not identical, adding two new phases to the ferroelectric nematic realm. The two I$_A$ phases exhibit similar, strongly peaked, diffuse x-ray scattering in the WAXS range ($1 < q < 2$ Å$^{-1}$) indicative of a distinctive mode of short-ranged, side-by-side molecular packing. The scattering of the I$_A$ phases at small $q$ is quite different in the two materials, however, with RM734 exhibiting a strong, single, diffuse peak at $q \sim 0.08$ Å$^{-1}$ indicating mesoscale modulation with ~80 Å periodicity, and DIO a sharper diffuse peak at $q \sim 0.27$ Å$^{-1} \sim (2\pi$/molecular length), with second and third harmonics, indicating that in the I$_A$ phase of DIO, short-ranged molecular positional correlation is smectic layer-like.

The mixtures were studied using standard liquid crystal phase analysis techniques, previously described [1,2,3,4,5] including depolarized transmission optical microscopic observation of LC textures and response to electric field, x-ray scattering (SAXS and WAXS), and techniques for measuring polarization and determining electro-optic response [4].

*Materials* – DIO *(*2,3´,4´,5´-tetrafluoro-[1,1´-biphenyl]-4-yl 2,6-difluoro-4-(5-propyl-1,3-dioxane-2-yl)benzoate, **Fig. S1**, compound *3*) is a rod-shaped molecule about 20 Å long and 5 Å in diameter, with a longitudinal electric dipole moment of about 11 Debye. DIO was first reported by Nishikawa et al. [6] and was synthesized by the Walba group as described in [4]. The synthesized compound was found to melt at $T$ = 173.6°C and in addition to a conventional nematic (N) phase, exhibited two additional mesogenic phases, a lamellar antiferroelectric (the SmZ$_A$) and the ferroelectric nematic (N$_F$). The transition temperatures on cooling were Iso – 173.6°C – N – 84.5°C – SmZ$_A$ – 68.8°C – N$_F$ – 34°C – X, very similar to the temperatures reported by Nishikawa.

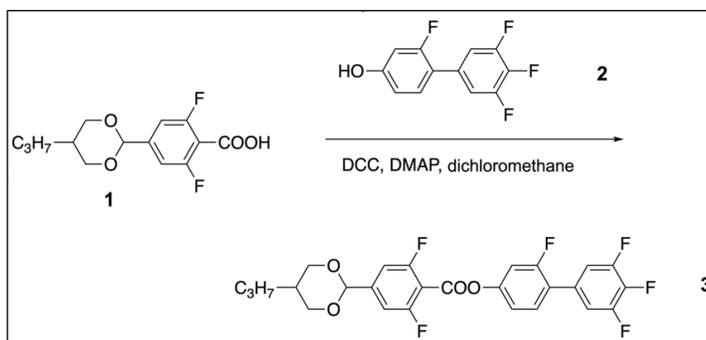

*Figure S1*: Synthesis scheme for DIO.

EMIM-TFSI {1-Ethyl-3-methylimidazolium bis(trifluoromethylsulfonyl)imide} was obtained from Millipore/Sigma and used without further purification.

*Methods – Obtaining the I$_A$$^{lam}$ phase in undoped DIO or DIO/IL mixtures* – Samples of DIO and its mixtures with ionic liquid were heated into the N phase at 120°C before they were loaded into capillaries or liquid crystal cells. After filling, the x-ray capillaries were quenched on a flat metal surface at $T$ = -19°C, by which means the entire volume of the doped LC rapidly transitioned to the optically transparent reentrant isotropic phase. No annealing was required. The capillaries were then heated back to room temperature to carry out x-ray diffraction in the dark phase, after which they were heated to the uniaxial nematic phase and cooled at -0.5°C per minute to selected temperatures where x-ray diffraction measurements of the N and SmZ$_A$ phases were carried out. In the EMIM/DIO mixtures, crystals of DIO typically started to appear when the sample was cooled to between 50°C and 40°C, making it challenging to obtain diffraction images of the N$_F$ phase. The capillary could subsequently be reheated to the nematic and quenched to the I$_A$ phase as many times as desired.

The phase transition temperatures in cells and capillaries were determined using depolarized transmission optical microscopy while cooling the samples at -0.5°C per minute. DIO crystals typically started to nucleate and grow at around 40°C in such temperature scans. In order to favor formation of the $I_A^{lam}$ phase rather than the crystal, it is necessary to cool the samples quickly, which prevents the crystal phase from growing to cover the whole volume before the transition to the $I_A^{lam}$ phase can take place. For example, crystallization could be suppressed by cooling at -3°C per minute, by which means a cell with the $I_A$ phase filling the entire volume could be obtained. In order to measure the precise temperature of the $N_F$ – $I_A$ transition, the cell was first quenched to 30°C and then cooled at -0.5°C per minute to the transition at about 27°C.

The $I_A^{lam}$ phase in the DIO/EMIM mixtures is apparently less thermodynamically stable than the crystal. At 20°C, the crystal phase was typically observed to start nucleating following overnight storage. However, the $I_A^{lam}$ phase was stable for at least five days if the sample was held at 9°C. Upon heating at 0.5°C per minute, the $I_A^{lam}$ phase transitioned to the $N_F$ phase at $T \sim 45$°C, after which crystals quickly grew in the sample.

*X-ray scattering* – For SAXS and WAXS experiments, LC samples were filled into thin-wall capillaries 0.7 to 1 mm in diameter. Data presented here are powder averages obtained on cooling using a Forvis microfocus SAXS/WAXS system with a photon energy of $CuK_\alpha$ 8.04 keV (wavelength = 1.54 Å). Each scan took ~1 hr at a given temperature.

*Polarized light microscopy* – Optical microscopy of LC cells viewed in transmission between crossed polarizer and analyzer, with such cells having the LC between uniformly spaced, surface-treated glass plates, provides key evidence for the macroscopic polar ordering, uniaxial optical textures, and fluid layer structure in LC phases and enables direct visualization of the director field, *n*(*r*), and, apart from its sign, of ***P***(***r***).

*Electro-optics* – For making electro-optical measurements, the mixtures were filled into planar-aligned, in-plane switching test cells (Instec, Inc.) with unidirectionally buffed alignment layers arranged antiparallel on the two plates, which were uniformly spaced 3.5 μm apart. In-plane ITO electrodes on one of the plates were spaced by a 1 mm wide gap and the buffing was along a direction rotated 3° from parallel to the electrode edges. Such surfaces give a quadrupolar alignment of the N and $SmZ_A$ directors along the buffing axis and polar alignment of the $N_F$ at each plate, leading to a director/polarization field in the $N_F$ phase that is parallel to the plates and has a π twist between the cell surfaces [3].

***Section S2*** – *SAXS & WAXS TEMPERATURE SCANS*

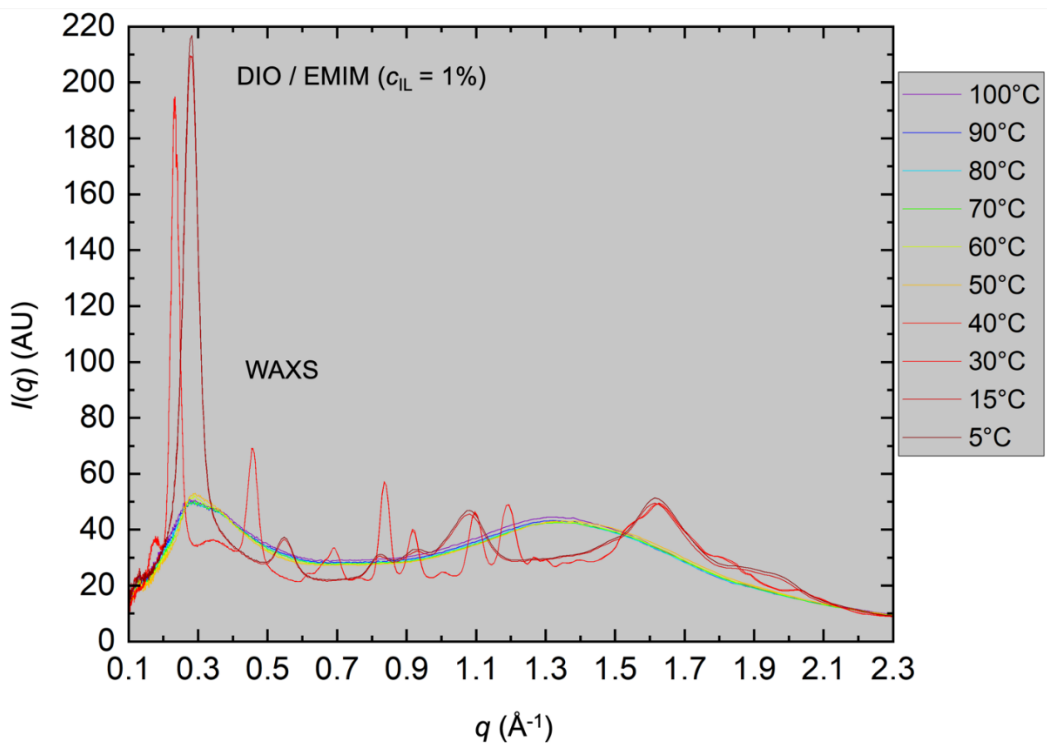

***Figure S2:*** WAXS scans vs. *T* of a DIO/EMIM ($c_{IL}$ = 1%) mixture. Diffraction measurements were made first in the $I_A^{lam}$ phase, on heating from 5°C to 15°C. The sample was then heated to the nematic phase at 100°C, and the remaining measurements were carried out on cooling to 30°C. The sample crystallized below 50°C.

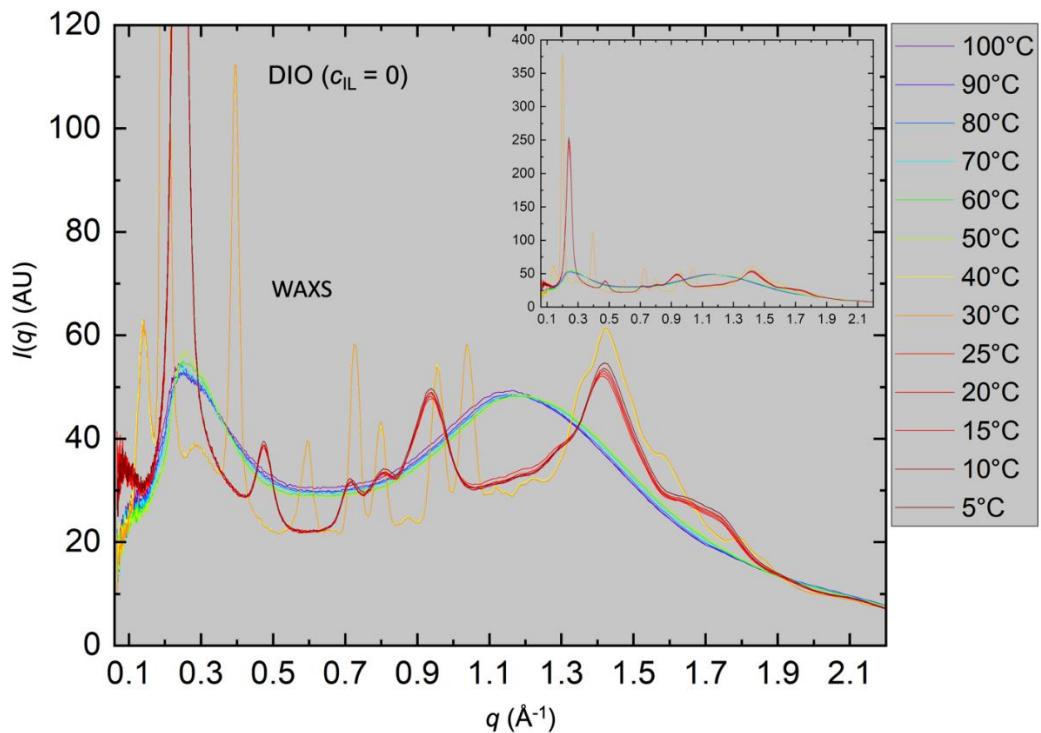

*Figure S3:* WAXS scans vs. *T* of undoped DIO. Diffraction measurements were made first in the $I_A^{lam}$ phase, on heating from 5°C to 25°C. The sample was then heated to the nematic phase at 100°C, and the remaining measurements were carried out on cooling to 30°C. The sample crystallized below 50°C.

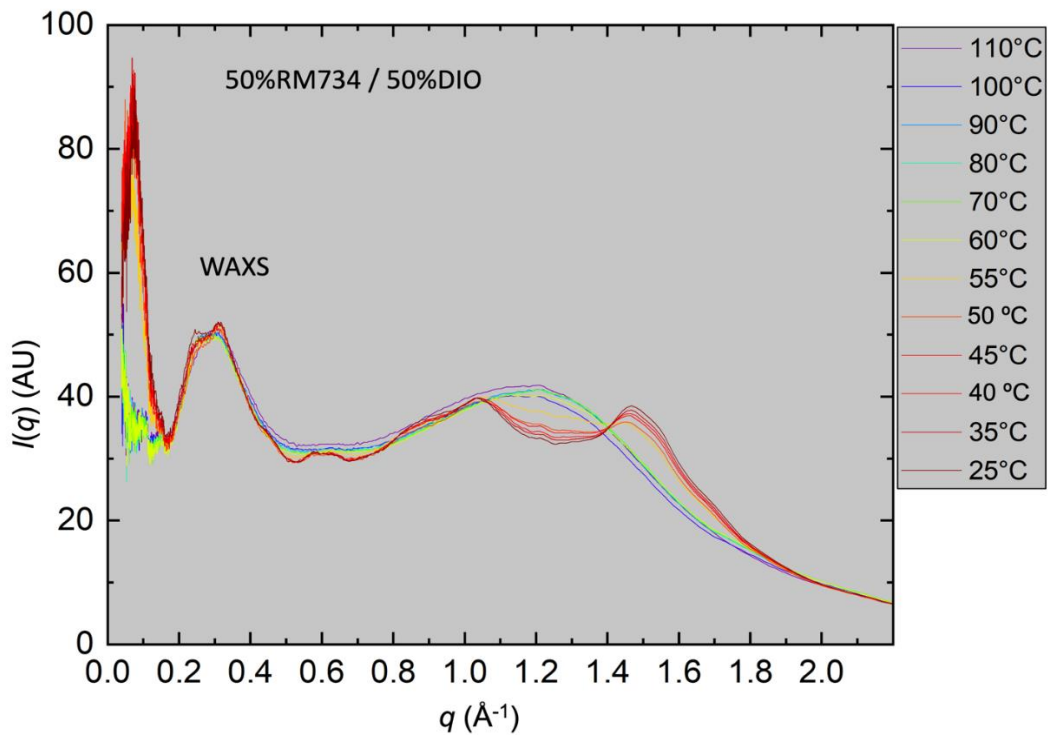

*Figure S4:* WAXS cooling scans vs. *T* of a 50% RM734/50% DIO mixture.

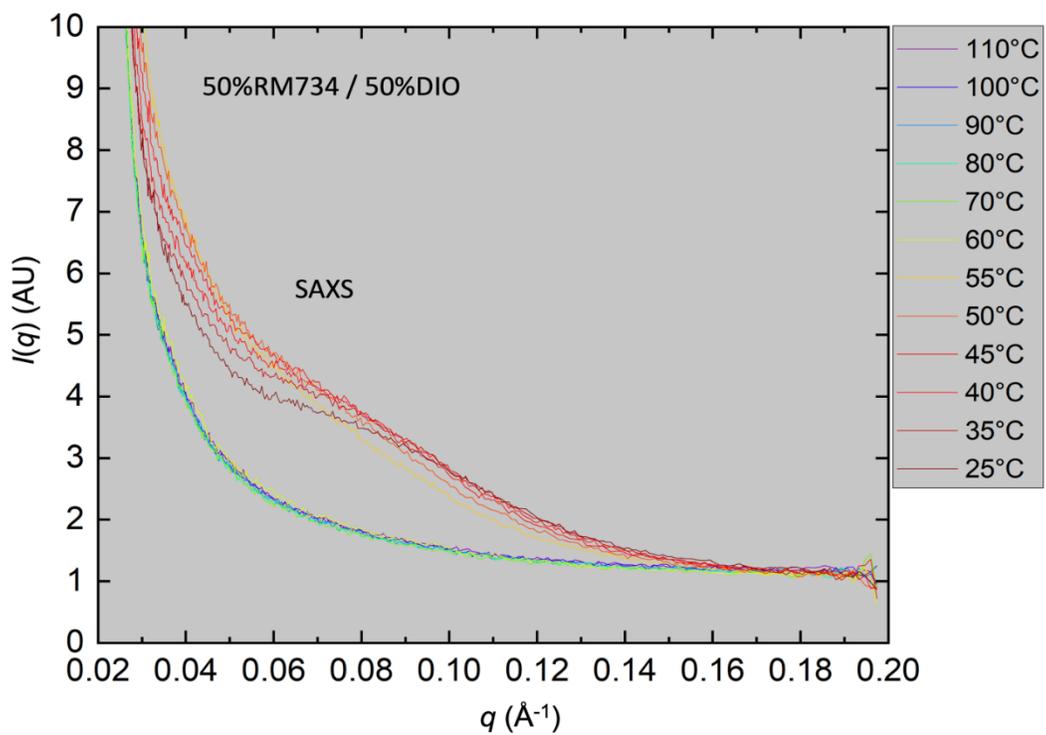

*Figure S5:* SAXS cooling scans vs. *T* of a 50% RM734/50% DIO mixture.

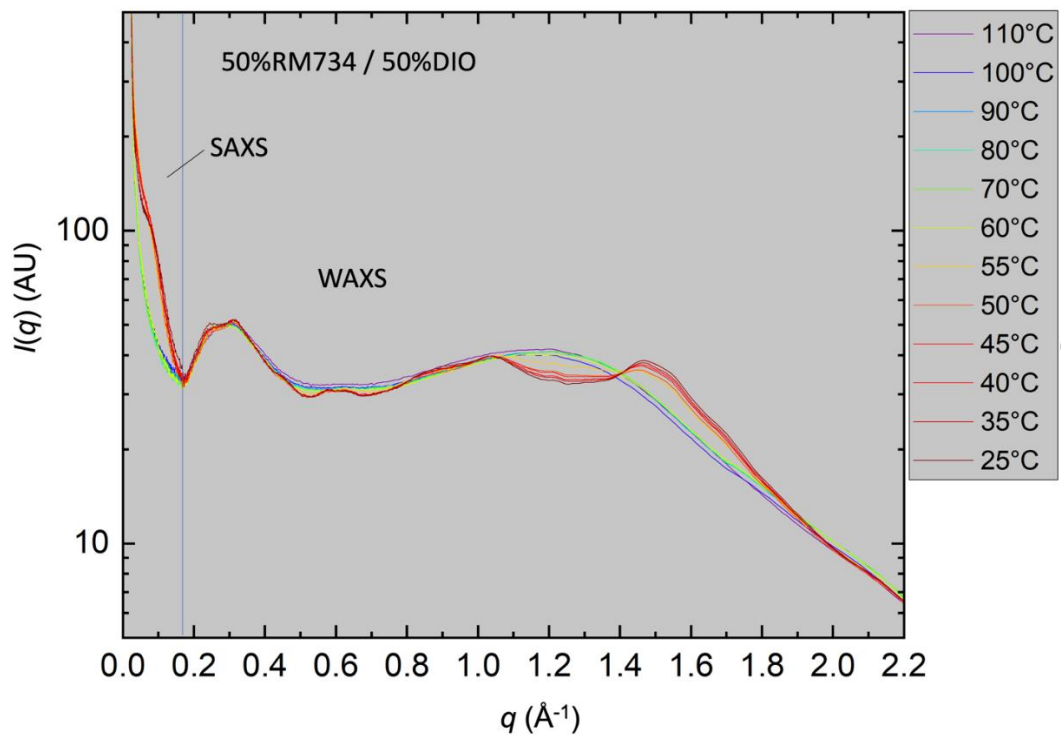

*Figure S6:* SAXS & WAXS cooling scans vs. *T* of a 50% RM734/50% DIO mixture.